# Can a Hurricane Be Managed?

Lawrence Sirovich, Brown University

Abstract

Based on realistic estimates of geophysical conditions it is demonstrated that by practical means; (1) a hurricane's intensity can be diminished before making landfall; (2) and other circumstances, potential hurricanes might be quenched before achieving critical strength.

Under the first strategy, cold deep ocean water cools the hurricane track to weaken the hurricane.

Analyses are facilitated by derivation of a novel exact solution that models a tropical depression, a mesoscale eddy, and the eye of a hurricane.

Under the second strategy, it is shown. the threat of a tropical depression can be diminished, in a timely manner.

Estimates of the power needed to perform timely ocean mixing show that this can be accomplished by high-performance submarines, and an exceptional *coefficient of performance*, $O(10^4)$.

The destructive power of a hurricane is functionally related to a hurricane's maximal wind speed, $V_m$. It is shown that a 20% reduction in maximal wind speed produces a year 50% reduction in destructive costs.

Complementary deliberations show the potential for producing rainfall in relief of drought.

Shown also is the opportunity to significantly modify vessel architecture to maximize the desired goals.

It is the contention of this paper that a practical framework exists to pursue means by which to reduce the tragedy and devastation caused by hurricanes.

1. Introduction

To paraphrase the description of another human calamity (Mukherjee 2010) , a hurricane (cyclone, typhoon) is the emperor of all meteorological disasters. With cyclonic diameters greater than a thousand kilometers, it is huge. With estimated energies of order $10^{19}$ Joules, equivalent to~100,000 medium-size atomic bombs (Monin 1972), it is a monster!

Hurricanes are fueled by radial inflow of an energetic ocean spray, collected at the sea surface, into the low-pressure core of the hurricane eye. This provides energy that escalates the cyclonically upward spiraling of the resulting intense atmospheric vortex. The overall process has been likened to a Carnot cycle (Emanuel 2003; Emanuel 1991). Beyond this, the atmospheric hurricane is meteorologically steered dynamically by an ambient atmosphere. A true depiction of hurricanes requires consideration of oceanography and meteorology interaction, (Pedlosky 2013). The present investigation explores methods which interfere with the fueling role of the ocean, in contrast to the high-profile, meteorological (seeding) attempts for altering hurricanes, of the last century, termed STORMFURY (Willoughby et al. 1985), that were deemed to be a failure. The presentation that follows addresses hurricane mitigation, is solely focused on oceanographic considerations; mention of the *atmospheric hurricane* will only be incidental.

Any attempt to modify the above described monster might seem foolhardy. Nevertheless, a hurricane on reaching landfall, is removed from its energy source, and undergoes a steady decrease in intensity. Hurricane intensity is measured by maximal hurricane velocity, $V_m$, and on reaching land is well modeled by (Kaplan and DeMaria 2001),

$$\frac{dV_m}{dt} = -\frac{V_m}{\tau}; \quad \tau \approx 10\,hr. \tag{1.1}$$

Thus, 10 hours after landfall, the strength of a hurricane falls by more than half. It is an empirical fact that a hurricane cannot form unless the sea surface temperature (SST) is greater than 26°C, (Gray 1979) . The possibility of cooling a portion of its track, in advance of hurricane arrival, is next investigated.

2 A Least Upper Bound for Work for Ocean Cooling

Typically, a warm ocean surface layer lies above a cold deep sea. The transition from the surface layer to the deep cold sea below is referred to as the thermocline (Pedlosky 2013), is depicted in figure 1 for the Caribbean, the Gulf of Mexico, and the North Atlantic (NOAA 2009).

The representative calculation will be performed for the North Atlantic case.

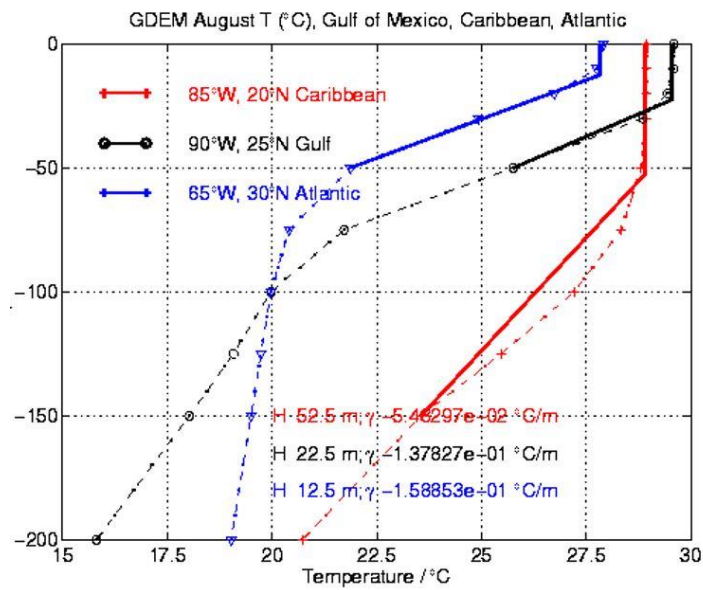

Figure 1. Three different thermocline locations as they typically appear in the month of August.

Consider a square meter cross-section column of seawater spanning the surface layer and thermocline, Figure 2, left. The aim is to mix the column, to obtain the lower temperature uniform column shown at the right. The argument is informal, based on reasonable approximations and estimates.

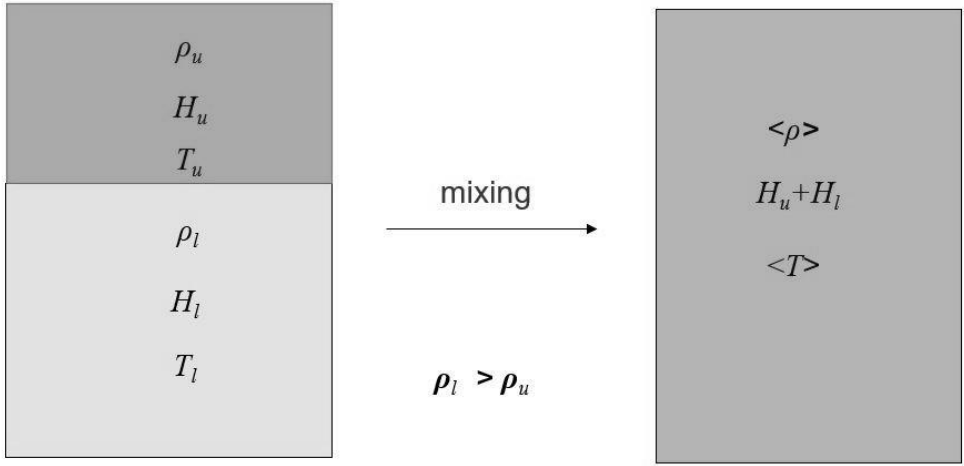

Figure 2. Left column: The case of the North Atlantic, with upper layer $H_u = 20$ m, and temperature $T_u = 27°C$; the lower layer $H_l = 50$ m, and at a temperature $T_l = 20°C$. Right column: represents the uniformly mixed upper and lower columns, and lower potential energy. Both columns are one m² cross-section.

For constant heat capacity, the temperature of the mixed state is considered for,

$$T_u - \langle T \rangle \approx 5°C, \tag{1.2}$$

a decrease greater than usually needed to reduce the surface layer below 26°C.

The difference in potential energies of the two columns of figure 2 represents the *minimal* needed work, $W$, to obtain the mixed state,

$$W/m^2 = (\rho_l - \rho_u) g \frac{H_u H_l}{2}, \tag{1.3}$$

which for $\rho_l = 1027 kg/m^3$ & $\rho_u = 1025 kg/m^3$ yields

$$W/m^2 = 10^4 \, Joules. \tag{1.4}$$

Emphasis on *minimal* for (1.4), since it is the **least upper bound** of required work. This provides a representative estimate and is analogous to the role played by a Carnot cycle in thermodynamics. As shown below it is an acceptable *ballpark* estimate of the true work needed for cooling.

Note that (1.4) is roughly the energy needed to illuminate a 200 W bulb for a minute. This key calculation, informs us that since,

$$\varepsilon = (\rho_l - \rho_u)/\rho_l \approx .2\%, \quad (1.5)$$

relatively little work is required for mixing. As discussed below an extremely high COP (coefficient of performance) is responsible for this outcome. Also see (Winters et al. 1995).

*Hurricane Mitigation*

(Gallacher, Rotunno, and Emanuel 1989) suggest that "a 2.5° C decrease in temperature near the core of the storm (hurricane) would suffice to shut down energy production entirely".

Nominal values for hurricane speed and and eye diameter are 20 km/h and 50 km., respectively. A reasonable guess for nuclear submarine speed, is ~67-83 km/h. From these estimates, and hurricane forecasting, it is certain that a submarine pack can intercept and in a timely manner lay down a carpet of cold ocean water to diminish the intensity of an oncoming cyclones. For example to create virtual landfall, 10 hours before true landfall, the track area of 50 km × 200 km ≈$10^{10}$ m$^2$ by (1.1) would require,

$$\overline{W} = 10^{14} \; Joules, \quad (1.6)$$

of energy to cool it by 5°C. While the extent of a hurricane might be 1000 km, it is fueled by an ocean area of diameter 50 km, a ratio of 1/20, which will figure in modeling estimates. Since submarine speeds are roughly four times hurricane speeds, forecast uncertainties become less critical.

As an example the Russian Shark class nuclear submarine, has a power rating of ≈$2\times10^8$ Joules/sec (Naval-Technology.com 2011). This is equivalent to the output of a power plant station of a small city, thus a nuclear submarine can be viewed as an ocean going power plant. For the 10 hours (=$3.6\times10^4$ sec.) duration needed to create the virtual landfall, this amounts to a total energy of ≈ $10^{13}$Joules. It follows from (1.6) that, roughly speaking, 10 submarines might be required to create the virtual landfall.

*Turbulent Mixing*

Cold sea water, raised from the depths, if released at the sea surface, falls back to its natural level, unless quickly mixed, say by turbulence, the most efficient mixer. Based on typical US nuclear submarine specifications (Virginia and Ohio class), a sub's beam is about 40 feet and the speed estimate ~67-83 km/h. Thus, a typical Reynolds number, *Re*, is

$$\text{Re} = O(10^8), \quad (1.7)$$

which implies a fully turbulent wake starting with a 40-foot stern.

*Hurricanes Costs*

Wind forces are proportional to $V_m^2$, however, hurricane damage is proportional to the rate of work, i.e., power, hence proportional to $V_m^3$. This key distinction suggests that if $V_m$ are diminished by 20%, costs are halved!

Estimated hurricane costs to world economies can vary from tens of billions to tens of trillions of dollars, depending on the criteria used in the studies (Kahn 2014; Mendelsohn and Saher 2011). Hence, reducing costs by half takes on profound economic significance.

*Coefficient of Performance*

Elementary thermodynamic arguments (Fermi 1956) show that for the nominal 50 km × 200 km ocean area, and a modest depth of 20 meters, to be cooled by 5°C, not by mixing, but by heat removal of a Carnot cycle, requires an energy,

$$dE \approx 4 \times 10^{18} J. \qquad (1.8)$$

On the other hand, the above deliberations accomplish this by making use of available deep cold water, lifted, and mixed with the warm surface water, compared with the work $\overline{W}$, (1.6). In effect the submarine pack act as a heat pump.

Equation(1.8) implies that the *coefficient of* performance is

$$COP_{cool} = dE / \overline{W} \approx 10,000. \qquad (1.9)$$

This is extraordinary compared to a *COP* of 2 or 3 for a conventional heat pump. At the heart of this energy leverage is the slimness of increase in ocean density with depth, (1.5).

*Improved Work Estimate*

The calculation of $\overline{W}$, (1.6), represents is the *minimal* required work. Elementary dimensional reasoning shows that the *true* work needed, $W_T$, has the functional form,

$$W_T / \overline{W} = f(\varepsilon, \text{Re}) \qquad (1.10)$$

where

$$\varepsilon = \frac{\rho_l - \rho_u}{\langle \rho \rangle} \left( \approx \frac{T_u - T_l}{\langle T \rangle} \right), \tag{1.11}$$

measures the gradient. It follows from (1.5) and (1.7) that (1.10) should be considered for $\varepsilon \downarrow 0, \text{Re} \uparrow \infty$ in which case (1.10) becomes

$$W_T \sim \varepsilon \times \overline{W}, \tag{1.12}$$

under a smoothness assumption on *f*. A more comprehensive analysis would include the Rossby number, *Ro* and the Froude number to account for Coriolis and gravitation affects. The former may be taken as infinite, and the latter might be included by redefining $\varepsilon = \varepsilon$ *(Fr)*, and thus a slight modification of (1.12).

A useful guide in these deliberations is the case of a passive scalar, e.g., a dye, in which case full mixing occurs, in the presence of turbulence, *without* additional work (Sreenivasan 1991). In view of (1.5), density differences are tiny, and as such are akin to a passive scalar, in which case mixing comes for *free*. This, and experience with dimensional reasoning suggests that (1.12) is unlikely to be off by more than a factor of 2. In the absence of experiment, this is the only support for estimates on the power needed for hurricane management.

*Submarine Modification*

There might be concern that the efficiency of the propulsion unit is not being factored into our deliberations. Since we are only looking for orders of magnitude, this point is a minor qualm. The propulsion unit is electric, and thus normally quite efficient. Of more importance is that submarine design is highly influenced by stealth demands, i.e., the need to avoid wake detection by satellite imaging. The present application is *free* of this restraint, and to the contrary a large wake is desirable, and thus efficiency considerations take on the different dimension.

For example, submarine modification might include a variable diameter propeller, possibly as large as the beam diameter of *Do*~40 feet, to enable the action of fully developed turbulence across the wake.

Wake growth, *D*, with distance downstream, *X*, is given by $D/Do = 1.25 \times (X/Do)^{.22}$, an empirical formula (MERRITT 1972). This predicts that after one sub length, ~150 m the wake diameter is ~33 m. Under this scenario, the work done in lifting the heavier deep ocean water is subsumed by turbulence.

*3.* Another Approach: Quelling of Tropical Depressions

Tropical depressions are storms of limited extent and strength, that are regarded as hurricane risks, and are routinely monitored by NOAA. An alternate strategy might be to dispatch submarines from well-chosen locations, with the mission of removing the potential storm threat. For example, hurricane Dorian, was recognized as a tropical depression, on August 23, 2019; a week later it exhibited cyclonic potential. To explore what might've been done in the intervening week, we remark that the ocean flow is dictated by vortex motion on a rotating sphere (Newton 2013).

The dynamics are governed by the Euler equations for a frame rotating with angular velocity, $\vec{\Omega}$, as will given by,

$$\rho \frac{d\vec{u}}{dt} + \nabla p = \rho \left( \nabla \frac{1}{2} |\vec{\Omega} \times \vec{r}|^2 - 2\vec{\Omega} \times \vec{u} \right), \tag{3.1}$$

(Kageyama and Hyodo 2006; Pedlosky 2013) where the 2 terms on the right-hand side represent the centripetal and Coriolis accelerations. For the earth's northern hemisphere $\vec{\Omega}$ is a vector pointing north, of magnitude

$$\Omega = 7.3 \times 10^{-5} \, rad \cdot s^{-1}. \tag{3.2}$$

The "Coriolis force" points rightward from the of flow direction $\vec{u}$; towards the *right bank* in the northern hemisphere.

To model the surface layer of the ocean, ignore vertical motion and consider the tangent plane z=0. This is given by the polar form of (3.1),

$$\begin{aligned} C &: \frac{\partial r u_r}{\partial r} + \frac{\partial u_\theta}{\partial \theta} = 0 \\ M_r &: \frac{\partial u_r}{\partial t} + u_r \frac{\partial u_r}{\partial r} + \frac{u_\theta}{r} \frac{\partial u_r}{\partial \theta} - \frac{u_\theta^2}{r} + \frac{1}{\rho} \frac{\partial p}{\partial r} = 2\Omega_o^2 r - 2\Omega_o u_\theta \\ M_\theta &: \frac{\partial u_\theta}{\partial t} + u_r \frac{\partial u_\theta}{\partial r} + \frac{u_\theta}{r} \frac{\partial u_\theta}{\partial \theta} + \frac{u_r u_\theta}{r} + \frac{1}{\rho r} \frac{\partial p}{\partial \theta} = 2\Omega_o u_r, \end{aligned} \tag{3.3}$$

where $\Omega_o = \Omega \sin \varphi$ is the local latitudinal rotation rate, in the absence of vertical motion.

Equations (3.3) have an exact cylindrically symmetric solution given by,

$$\begin{aligned} u_\theta &= \Omega_o r + \beta / r, \\ u_r &= \alpha / r, \\ \frac{1}{\rho} \frac{\partial p}{\partial r} &= -\left( \frac{\partial}{\partial r} (u_r^2 / 2) - \frac{u_\theta^2}{r} \right). \end{aligned} \tag{3.4}$$

The first term of $u_\theta$ is the relevant uniform rotation and *(α, β)*, of units $\ell^2 / t$, are *source* strengths, to be discussed below.

As an illustration suppose $\beta = 0$, then streamlines correspond to a *source*, at the origin, and the curvature of the streamlines due to the Coriolis acceleration. The stream function, from (3.4) in dimensionless form, is given by

$$\psi = \alpha\theta - \frac{\Omega_o r^2}{2}. \tag{3.5}$$

An exemplar of the stream function (3.5) is shown in the following figure.

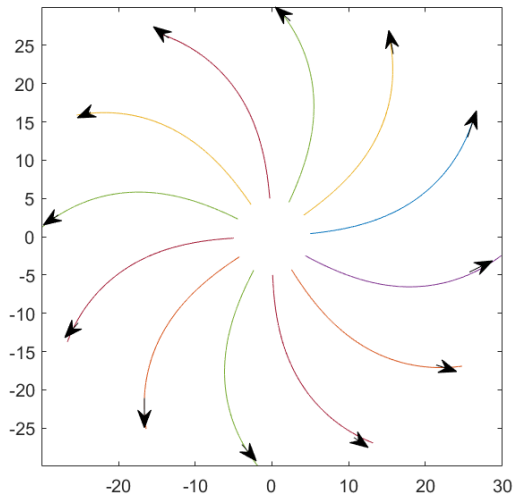

Figure 3. Streamlines for the inviscid model of a tropical depression, based on a circular region around the origin, corresponding to a radius of 30 miles The streamlines correspond to $\theta = 0°$ to $\theta = 360°$. The streamlines shown above model a tropical depression at a nominal North Atlantic latitude. More southerly latitudes have more radial streamlines, more northerly latitudes the streamlines show more swirl.

For practical application, a NOAA snapshot of a tropical depression, at some location at which the rotation rate $\Omega_o$ is known, is fit to (3.4), along with a determination of where, in the data, **r** = 0. Thus, data furnishes $\alpha$, $\beta$, and **r** hence an analytical *shape* is conferred on the tropical depression, and its center.

 In keeping with the general theme, to inhibit the cyclonic development by cooling, the surface layer should be subjected to mixing, from a large depth of ocean, ~70 m. Since a tropical depression is small ~$O(10^2)$ *km,* only a small region, say of diameter 20 km, around the now known center, as suggested earlier, a small relative area need be mixed, and thus few submarines are required. Another scheme is to have the submarine pack execute a circular, cyclonic band around the origin solely in the surface layer, hence without mixing. This confers an additional circulation on the tropical depression, with an accompanying Coriolis *force*, which can lift or force down the surface layer, depending on the signature of the circulation.

It is worth mentioning in passing that *Easterlies*, aerodynamically, can steer the atmospheric storm system northward or southward. For squelching a tropical depression northward is desirable, since disturbances north of the 20[th] latitude rarely develop into cyclones (Knaff et al.

2013; Knaff, Longmore, and Molenar 2014). At more northerly latitudes the surface layer becomes cooler, and greater Coriolis force pumps deeper water to the surface.

This strategy, since relatively brief, also diminishes moisture accumulation, hence even if the storm is not squelched less rainfall accompanies the hurricane.

Note that a reversal of the above simple reasoning leads to a method which might enhance the hurricane initiation, for the purpose of increasing rainfall.

4. Two Related Applications

The above source-rotational model, (3.4) also models two other phenomena, not unrelated to hurricane study, namely the eye of a hurricane, and mesoscale eddies.

*The Eye of the Hurricane*

In this case assume that the diameter of the hurricane eye is known, $D = 2 \times R$, as is $V_m$. It then follows that the rotating system has an angular velocity given by,

$$\Omega_H = V_m / (2\pi R). \tag{3.6}$$

A simple model would be,

$$\begin{aligned} u_\theta &= \Omega_H \times r, 0 < r < R \\ u_\theta &= \Omega_H \times R / r, R < r. \end{aligned} \tag{3.7}$$

More generally, if detailed data is available for the eye of the hurricane, then (3.4) could be fit by a least squares procedure to determine the constants $(\alpha, \beta)$. The induced flow produces an outward Coriolis force, hence an outward radial flow. Physically, this signifies an outflow from the origin, of cold deep water lifted to the surface layer, a negative feedback effect, since it cools the surface layer. Evidence of this effect is seen in the imaging of hurricane Gilbert (NOAA 2009) that left a trail of cold deep ocean water lasting weeks after its passing through the Gulf.

*Mesoscale Eddies*

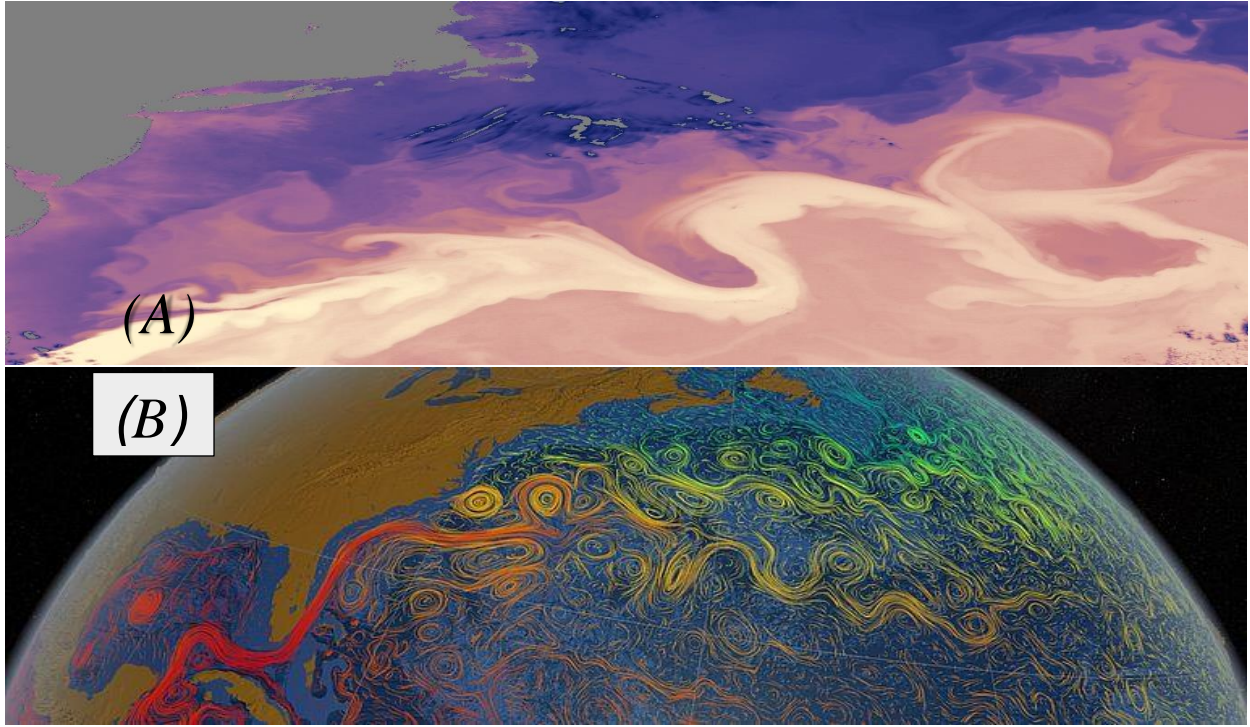

Figure 4 (A) Mesoscale eddy formation. (B) As cast off the Gulf Stream, in a von Karman vortex trail.

The model also can be applied to the large (100 – 200 km), mesoscale, eddies, Figure 4, as for example spawned off of the Gulf Stream (Chelton et al. 2007), that are remotely sensed and monitored. The data in the cited reference may be used to fit individually measured eddies again by a least squares fitting of (3.4).

The eddies shown in figure 4(A) form unstable loops that pinch off the Gulf Stream. Clearly, these show clockwise preference for a northern loop, that captures cold ocean, and that have counterclockwise preference for a southern loop, that captures warm ocean. The fact that the former are referred to as *warm*, and *cold*, respectively, is in agreement with the above description of the model, and adds validity to the contention that (3.4)models mesoscale eddies.

Also note that "warm" eddies persist for relatively long times, even though the governing equations contain a counterclockwise preference. The persistence of such eddies have been verified by remote-sensing observations. Mesoscale eddies have been linked to hurricane will development (Ma et al. 2017), and might be made to play a role cyclonic initiation and annihilation.

7. Remarks

The answer to how long a cooled ocean cold persists, is ironically, answered by satellite imagery of hurricane Gilbert (NOAA 1988). On September 14, 1988, Gilbert moved across the northwest coast of the Yucatán Peninsula, and in so doing, churned to the surface, cold deep waters, so that a wide swath of ocean fell from 28°C to 24°C. The cited imagery showed this to still be the case five days later. More generally, (Knaff et al. 2013), report that the cooled hurricane wake persists for weeks.

Atmospheric hurricanes are steered by ambient meteorological conditions. Therefore, to combat drought one might look for circumstances where enhancing hurricane initiation might succeed. California needs rain, but rarely experiences hurricanes. A search for opportunities, would likely change the odds. A more likely opportunity for hurricane enhancement, is New South Wales, Australia, which does have a history of hurricanes, and is presently experiencing a devastating drought. There should be no fear of a breakaway hurricane since we have presented the tools for harnessing such a situation, without losing the rainfall.

The aim of this paper has been to provide examples, estimates, and frameworks for mitigating and preventing the economic devastation caused by cyclones. It is the contention of the author that although no *proof of concept* has been demonstrated, there is compelling evidence that the proposed framework be put to the test. The extreme nature of the physical problem puts computer and experimental modeling out of reach. Modification of a single traditional submarine, along with a test program that includes satellite imagery would provide a relatively low-cost answers to many questions. A test of the strategy of quenching potential disturbances, could be undertaken almost immediately.

Initiation/suppression of a cyclonic dynamics is a critical event, and as such involves a relatively small energetic range. This suggests that the actual ignition and retardation of a hurricane may be controlled by modest changes to the state of a tropical depression, an observation that adds to the feasibility of the proposed hurricane management program.

The costs of a full-scale testing program might appear to be immense, but are flight specks compared to the estimated $1 trillion cost of cyclones over the next 10 years.. Any risk reward calculation makes pursuit of this project compelling.

Acknowledgments. Thanks to many friends and acquaintances for their encouraging remarks. Helpful observations by E. Meiburg, J. Pedlosky, P. Newton and K. R. Sreenivasan, found a place in the manuscript.

Declaration of Interests: The author is unaware of any conflict of interest.